\documentclass{llncs}
\usepackage{epsfig}
\usepackage{subfigure}

\usepackage{amsmath}
\usepackage{color}
\usepackage{times}
\usepackage{amsfonts}
\usepackage{subfigure}
\usepackage{epsfig}
\usepackage[english]{babel}
\usepackage{graphicx}
\usepackage{hyperref}
 \usepackage{verbatim}
\newcommand{\gcc}{{\tt 403.gcc}~}
\newcommand{\sphinx}{{\tt 482.sphinx}~}
\newcommand{\row}{{\tt row\_major}~}
\newcommand{\col}{{\tt col\_major}~}
\newcommand{\nerr}{nRMSE~}
\newcommand{\err}{RMSE~}

\title{On the importance of nonlinear modeling in computer performance prediction}
\author{Joshua   Garland \and Elizabeth Bradley}

\institute{University of Colorado, Boulder CO 80309-0430, USA}

\begin{document}

\maketitle

\begin{abstract}

Computers are nonlinear dynamical systems that exhibit complex and
sometimes even chaotic behavior.  The models used in the computer
systems community, however, are linear.  This paper is an exploration
of that disconnect: when linear models are adequate for predicting
computer performance and when they are not.  Specifically, we build
linear and nonlinear models of the processor load of an Intel i7-based
computer as it executes a range of different programs.  We then use
those models to predict the processor loads forward in time and
compare those forecasts to the true continuations of the time series.

\end{abstract}

\section{Introduction}

Accurate prediction is important in any number of applications.  In a
modern multi-core computer, for instance, an accurate forecast of
processor load could be used by the operating system to balance the
workload across the cores.  The traditional models used in the
computer systems community use linear, time-invariant (and often
stochastic) methods, e.g., autoregressive moving average (ARMA),
multiple linear regression, etc. \cite{jain-artof}.  While these
models are widely accepted---and for the most part easy to
construct---they cannot capture the nonlinear interactions that have
recently been shown to play critical roles in a computer's
performance~\cite{berry06,mytkowicz09}.  As computers become more and
more complex, these interactions are beginning to cause
problems---e.g., hardware design ``improvements'' that do not work as
expected.  Awareness about this issue is growing in the computer
systems community~\cite{tippdirk}, but the modeling strategies used in
that field have not yet caught up with those concerns.

An alternative approach that captures those complex effects is to
model a computer as a nonlinear dynamical
system~\cite{mytkowicz09,todd-phd}---or as a {\sl collection} of
nonlinear dynamical systems, i.e., an iterated function system
\cite{zach-ifs-Chaos}.  In this view, the register and memory contents
are treated as state variables of these dynamical systems.  The logic
hardwired into the computer, combined with the code that is executing
on that hardware, defines the system's dynamics---that is, how its
state variables change after each processor cycle.  As described in
previous IDA papers \cite{zach-IDA10,josh-ida2011}, this framework
lets us bring to bear the powerful machinery of nonlinear time-series
analysis on the problem of modeling and predicting those dynamics.  In
particular, the technique called {\sl delay-coordinate embedding} lets
one reconstruct the state-space dynamics of the system from a
time-series measurement of a single state
variable\footnote{Technically, the measurement need only be a smooth
  function of at least one state variable}.  One can then build
effective prediction models in this embedded space.  One of the first
uses of this approach was to predict the future path of a ball on a
roulette wheel, as chronicled in~\cite{pie} and revisited recently
in~\cite{small12}.  Nonlinear modeling and forecasting methods that
rely on delay-coordinate embedding have since been used to predict
signals ranging from currency exchange rates to Bach fugues;
see~\cite{casdagli-eubank92,weigend-book} for good reviews.
\label{page:roulette}

This paper is a comparison of how well those two modeling
approaches---linear and nonlinear---perform in a classic computer
performance application: forecasting the processor loads on a CPU.  We
ran a variety of programs on an Intel i7-based machine, ranging from
simple matrix operation loops to SPEC cpu2006 benchmarks.  We measured
various performance metrics during those runs: cache misses, processor
loads, branch-prediction success, and so on.  The experimental setup
used to gather these data is described in Section~\ref{sec:methods}.
From each of the resulting time-series data sets, we built two models:
a garden-variety linear one (multiple linear regression) and a basic
nonlinear one: the ``Lorenz method of analogues,'' which is
essentially nearest-neighbor prediction in the embedded state space
\cite{lorenz-analogues}.  Details on these modeling procedures are
covered in Section~\ref{sec:models}.  We evaluated each model by
comparing its forecast to the true continuation of the associated time
series; results of these experiments are covered in
Section~\ref{sec:prediction}, along with some explanation about when
and why these different models are differently effective.  In
Section~\ref{sec:conclusion}, we discuss some future directions and
conclude.

\section{Experimental Methods}\label{sec:methods}

The testbed for these experiments was an HP Pavilion Elite computer
with an Intel Core\textsuperscript{\textregistered} i7-2600 CPU
running the 2.6.38-8 Linux kernel.  This so-called ``Nehalem'' chip is
representative of modern CPUs; it has eight cores running at 3.40Ghz
and an 8192 kB cache.  Its kernel software allows the user to monitor
events on the chip, as well as to control which core executes each
thread of computation.  This provides a variety of interesting
opportunities for model-based control.  An effective prediction of the
cache-miss rate of individual threads, for instance, could be used to
preemptively migrate threads that are bogged down waiting for main
memory to a lower-speed core, where they can spin their wheels without
burning up a lot of power\footnote{Kernels and operating systems do
  some of this kind of reallocation, of course, but they do so using
  current observations (e.g., if a thread is halting ``a lot'') and/or
  using simple heuristics that are based on computer systems knowledge
  (e.g., locality of reference).}.

To build models of this system, we instrumented the kernel software to
capture performance traces of various important internal events on the
chip.  These traces are recorded from the hardware performance
monitors (HPMs), specialty registers that are built into most modern
CPUs in order to log hardware event information.  We used the {\tt
  libpfm4} library, via PAPI \cite{papi}, to interrupt the executables
periodically and read the contents of the HPMs.  At the end of the
run, this measurement infrastructure outputs the results in the form
of a time series.  In any experiment, one must be attentive to the
possibility that the act of measurement perturbs the dynamics under
study.  For that reason, we varied the rate at which we interrupted
the executables, compared the results, and used that comparison to
establish a sample rate that produced a smooth measurement of the
underlying system dynamics.  A detailed explanation of the mechanics
of this measurement process can be found in
\cite{zach-IDA10,mytkowicz09,todd-phd}.
%
%

\begin{figure}
  \centering
  \includegraphics[width=0.49\columnwidth]{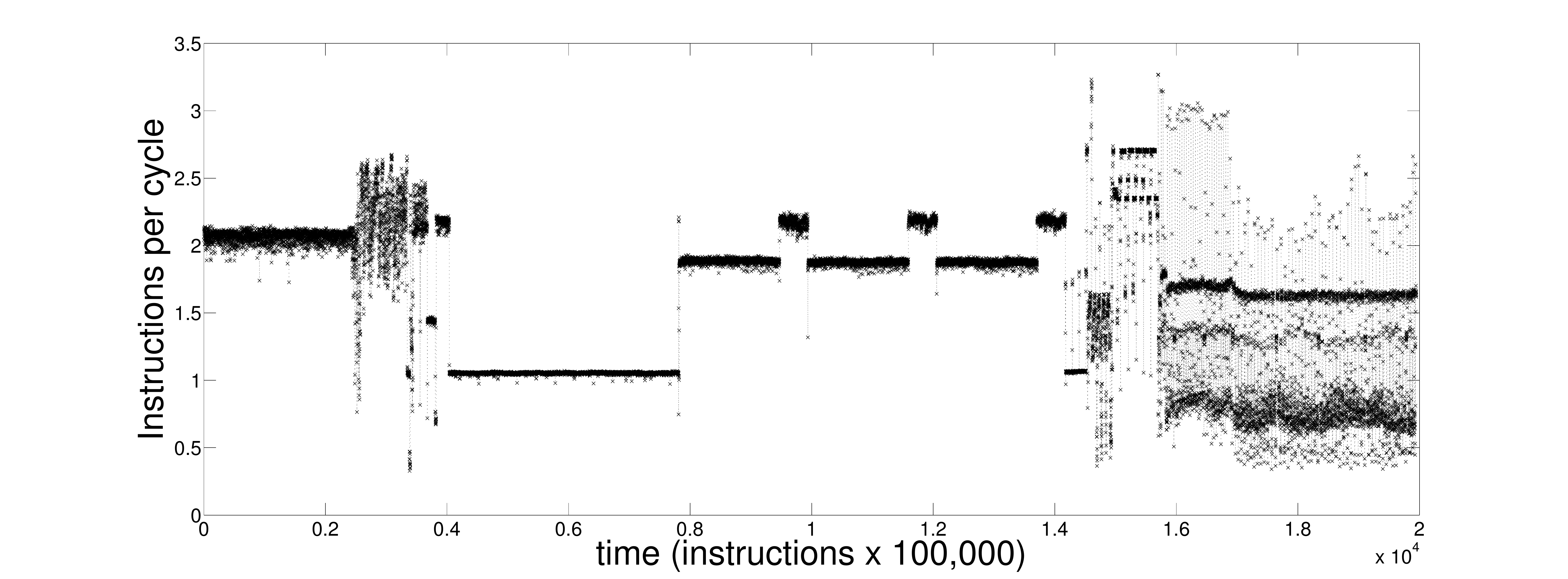}
  \includegraphics[width=0.49\columnwidth]{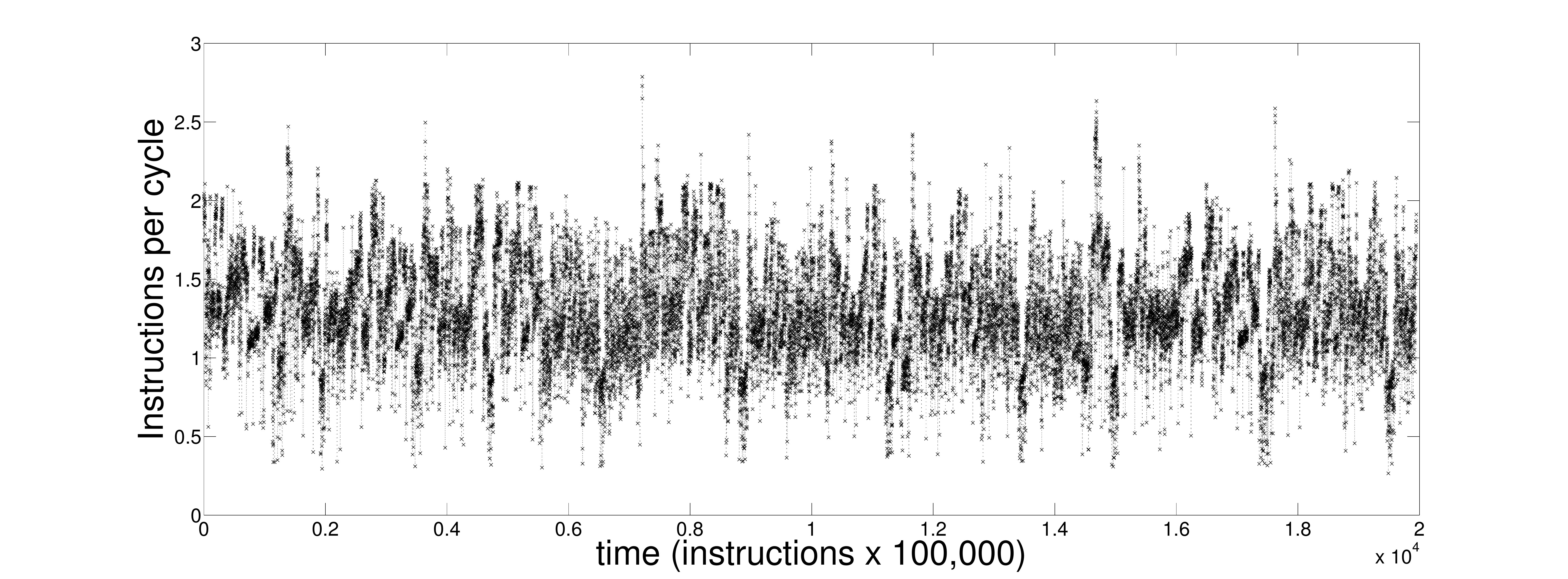}

{\sphinx \hspace*{1.5truein} \gcc}

  \caption{Processor load traces of the programs studied here}
  \label{fig:ipctrace}
\end{figure}

The dynamics of a running computer depend on both hardware and
software.  We ran experiments with four different C programs: two
benchmarks from the SPEC cpu2006 benchmark suite (the \gcc compiler
and the \sphinx speech recognition system) and two four-line programs
(\col and \row)
\label{page:programs}
that repeatedly initialize a matrix---in column-major and row-major
order, respectively.  These choices were intended to explore the range
of current applications.  The two SPEC benchmarks are complex pieces
of code, while the simple loops are representative of repetitive
numerical applications.  \gcc works primarily with integers, while
\sphinx is a floating-point benchmark.  Row-major matrix
initialization works naturally with modern cache design, whereas the
memory accesses in the \col loop are a serious challenge to that
design, so we expected some major differences in the behavior of these
two simple loops.  
Figure~\ref{fig:ipctrace} shows traces of the instructions executed
per cycle, as a function of time, during the execution of the two SPEC
benchmarks on the computer described in the first paragraph of this
section.  There are clear patterns in the processor load during the
operation of \sphinx.  During the first 250 million instructions of
this program's execution, roughly two instructions are being carried
out every cycle, on the average, by the Nehalem's eight cores.
Following that period, the processor loads oscillate, then stabilize
at an average of one instruction per cycle for the period from 400-800
million instructions.  Through the rest of the trace, the dynamics
move between different regimes, each with characteristics that reflect
how well the different code segments can be effectively executed
across the cores.
\label{page:sphinx-stochastic}
The processor load during the execution of \gcc, on the other hand,
appears to be largely stochastic.  This benchmark takes in and compiles
a large number of small C files, which involves repeating a similar
process, so the lack of clear regimes makes sense.
The dynamics of \row and \col (not shown due to space constraints)
were largely as expected.  The computer cannot execute as many
instructions during \col because of the mismatch between its
memory-access pattern and the design of the cache, so the baseline
level of the \col trace is much lower than \row.  Temporally, the \row
trace looks very much like \gcc: largely stochastic.  \col, on the
other hand, has a square-wave pattern because of the periodic stalls
that occur when it requests data that are not in the cache.

The following section describes the techniques that we use to build
models of this time-series data.

\section{Modeling computer performance data}
\label{sec:models}

\subsection{Overview}

The goal of this paper is to explore the effectiveness of linear and
nonlinear models of computer performance.  Many types of models, of
both varieties, have been developed by the various communities that
are interested in data analysis.  We have chosen multiple linear
regression models as our {\sl linear} exemplar because that is the
gold standard in the computer systems literature~\cite{tippdirk}.  In
order to keep the comparison as fair as possible, we chose the Lorenz
method of analogues, which is the simplest of the many models used in
the nonlinear time-series analysis community, as our {\sl nonlinear}
exemplar.  In the remainder of this section, we give short overviews
of each of these methods; Sections~\ref{sec:multilinear}
and~\ref{sec:nonlinear} present the details of the model-building
processes.

\subsubsection{Multiple Linear Regression Models}
\label{sec:multilinear-overview}

Suppose we have a set of $n$ observations from a system, where the $i^{th}$ observation 
includes $m+1$ measurements: a scalar response variable $r_i$ and a
vector of $m$ explanatory variables $[e_{i1},\dots,e_{im}]$.
Working from those $n$ observations
$\{([e_{i1},\dots,e_{im}],r_i)\}_{i=1}^n$, multiple linear regression
(MLR) models the response variable via a linear combination of the
explanatory variables---that is:
%
\begin{equation}\label{eq:hyper-plane}
r_i = [1, e_{i1},..., e_{im}]\vec{\beta}
\end{equation}
where $\vec{\beta} = [\beta_1,\dots,\beta_{m+1}]^T$ is a vector of
$m+1$ fit parameters \cite{jain-artof,regression}.  We estimated
$\vec{\beta}$ by using ordinary least squares to minimize the sum of
squared residuals (SSR) between the set of observations and the set of
hyperplanes defined in equation (\ref{eq:hyper-plane}):
$$SSR(\vec{\beta}) = \sum_{i=1}^{n}(r_i -[1, e_{i1},..., e_{im}]\vec{\beta})^2 = (\vec{r}-E\vec{\beta})^T(\vec{r}-E\vec{\beta})$$
Here, $E$ is a $n$ by $(m+1)$ matrix, with rows $\{[1, e_{i1},...,
e_{im}]\}_{i=1}^{n}$.  The $\vec{\beta}$ that minimizes this sum,
$\hat{\beta} = (E^TE)^{-1}E^T\vec{r}$, is called the {\sl
ordinary-least-squares estimator} of $\vec{\beta}$ \cite{regression}.

In practice, it is customary to measure everything that might be an
explanatory variable and then use a method called stepwise backward
elimination to reduce the model down to what is actually important.
And in order to truly trust an MLR model, one must also verify that
the data satisfy some important assumptions.  These procedures are
discussed in more detail in Section~\ref{sec:multilinear}.

\begin{quote}

{\sl Advantages:} MLR models are simple and easy to construct and use,
as they only require a small vector-vector multiplication at each
prediction step.

\smallskip

{\sl Disadvantages:} strictly speaking, MLR can only be used to model
a linear deterministic system that is measured without any error.  MLR
models involve multiple explanatory variables and cannot predict more
than one step ahead.
\end{quote}

 \subsubsection{Nonlinear Models}
\label{sec:nonlinear-overview}

Delay-coordinate embedding \cite{packard80,sauer91,takens} allows one
to reconstruct a system's full state-space dynamics from time-series
data like traces in Figure~\ref{fig:ipctrace}.  There are only a few
requirements for this to work.  The data, $x_i(t)$, must be evenly
sampled in time ($t$) and both the underlying dynamics and the
measurement function---the mapping from the unknown $d$-dimensional
state vector $\vec{Y}$ to the scalar value $x_i$ that one is
measuring---must be smooth and generic.  When these conditions hold,
the delay-coordinate map
\begin{equation}\label{eqn:takens}
F(\tau,d_{embed})(x_i) = ([x_i(t), ~ x_i(t+\tau), ~ \dots , ~x_i(t+d_{embed}\tau)])
\end{equation}
from a $d$-dimensional smooth compact manifold $M$ to
$\mathbb{R}^{2d+1}$ is a diffeomorphism on $M$ \cite{sauer91,takens}:
in other words, that the reconstructed dynamics and the true (hidden)
dynamics have the same topology.  This method has two free parameters,
the delay $\tau$ and the embedding dimension $d_{embed}$, which must
also meet some conditions for the theorems to hold, as described in
Section~\ref{sec:nonlinear}.  Informally, delay-coordinate embedding
works because of the internal coupling in the system---e.g., the fact
that the CPU cannot perform a computation until the values of its
operands have been fetched from some level of the computer's memory.
This coupling causes changes in one state variable to percolate across
other state variables in the system.  Delay-coordinate embedding is
designed to bring out those indirect effects explicitly and
geometrically.

The mathematical similarity of the true and reconstructed dynamics is
an extremely powerful result because it guarantees that $F$ is a good
model of the system.  As described on page~\pageref{page:roulette},
the nonlinear dynamics community has recognized and exploited the
predictive potential of these models for some time.  Lorenz's method
of analogues, for instance, is essentially nearest-neighbor prediction
in the embedded space: given a point, one looks for its nearest
neighbor and then uses {\sl that} point's future path as the
forecast~\cite{lorenz-analogues}.  Since computers are deterministic
dynamical systems~\cite{mytkowicz09}, these methods are an effective
way to predict their performance.  That claim, which was first made
in~\cite{josh-ida2011}, was the catalyst for this paper---and the
motivation for comparison of linear and nonlinear models that appears
in the following sections.

\begin{quote}

{\sl Advantages:} models based on delay-coordinate embeddings capture
nonlinear dynamics and interactions, which the linear models ignore,
and they can be used to predict forward in time to arbitrary horizons.
They only require measurement of a single variable.

\smallskip
{\sl Disadvantages:} these models are more difficult to construct, as
estimating good values for their two free parameters can be quite
challenging.  The prediction process involves near-neighbor
calculations, which are computationally expensive.
 
\end{quote}

\subsection{Building MLR forecast models for computer performance traces}\label{sec:multilinear}

In the experiments reported here, the response variable is the
instructions per cycle (IPC) executed by the CPU.
Following~\cite{tippdirk}, we chose the following candidate
explanatory variables {\sl (i)} instructions retired {\sl (ii)} total
L2 cache\footnote{Modern CPUs have many levels of data and instruction
  caches: small, fast memories that are easy for the processor to
  access.  A key element of computer design is anticipating what to
  ``fetch'' into those caches.}  misses {\sl (iii)} number of branches
taken {\sl (iv)} total L2 instruction cache misses {\sl (v)} total L2
instruction cache hits and {\sl (vi)} total missed branch predictions.
The first step in building an MLR model is to ``reduce'' this list:
that is, to identify any explanatory variables---aka {\sl
  factors}---that are meaningless or redundant.  This is important
because unnecessary factors can add noise, obscure important effects,
and increase the runtime and memory demands of the modeling algorithm.

We employed the stepwise backward elimination method~\cite{Faraway},
with the threshold value (0.05) suggested in~\cite{jain-artof}, to
select meaningful factors.  This technique starts with a ``full
model''---one that incorporates every possible factor---and then
iterates the following steps:
\begin{enumerate}
\item If the $p$-value of any factor is higher than the threshold,
  remove the factor with the largest $p$-value
\item Refit the MLR model
\item If all $p$-values are less than the threshold, stop; otherwise
  go back to step 1
\end{enumerate}
For all four of the traces studied here, this reduction algorithm
converged to a model with three factors: L2 total cache misses, number
of branches taken, and L2 instruction cache misses.

MLR models are meant to {\sl explain} the value of the response
variable in terms of the values of the explanatory variables, but they
can also be used to {\sl predict} it.  To do this, one takes a
measurement of each of the factors that appear in the reduced model
(say, $[e_1,\dots,e_m]$).  One then makes a prediction of IPC by
simply evaluating the function $[1,e_1,\dots,e_m]\hat{\vec{\beta}}$
and assigning that response to the \emph{next} time-step, i.e,
$r_{t+1}$.  That is how the predictions in the next section were
constructed.


Like any model, MLR is technically only valid if the data meet certain
conditions.  Two of those conditions are {\sl not} true for
computer-performance traces: linear relationship between explanatory
and response variables (which was disproved in~\cite{mytkowicz09}) and
normal distribution of errors, which is clearly not the case in our
data, given the nonlinear trends in residual quantile-quantile plots
of our data (not shown).  Despite this, MLR models are used routinely
in the computer systems community~\cite{tippdirk}.  And they actually
work surprisingly well, as the results in Section~\ref{sec:prediction} show.

\subsection{Building nonlinear forecast models for computer performance traces}
\label{sec:nonlinear}

\begin{figure}
  \centering
  \includegraphics[width=.49\columnwidth]{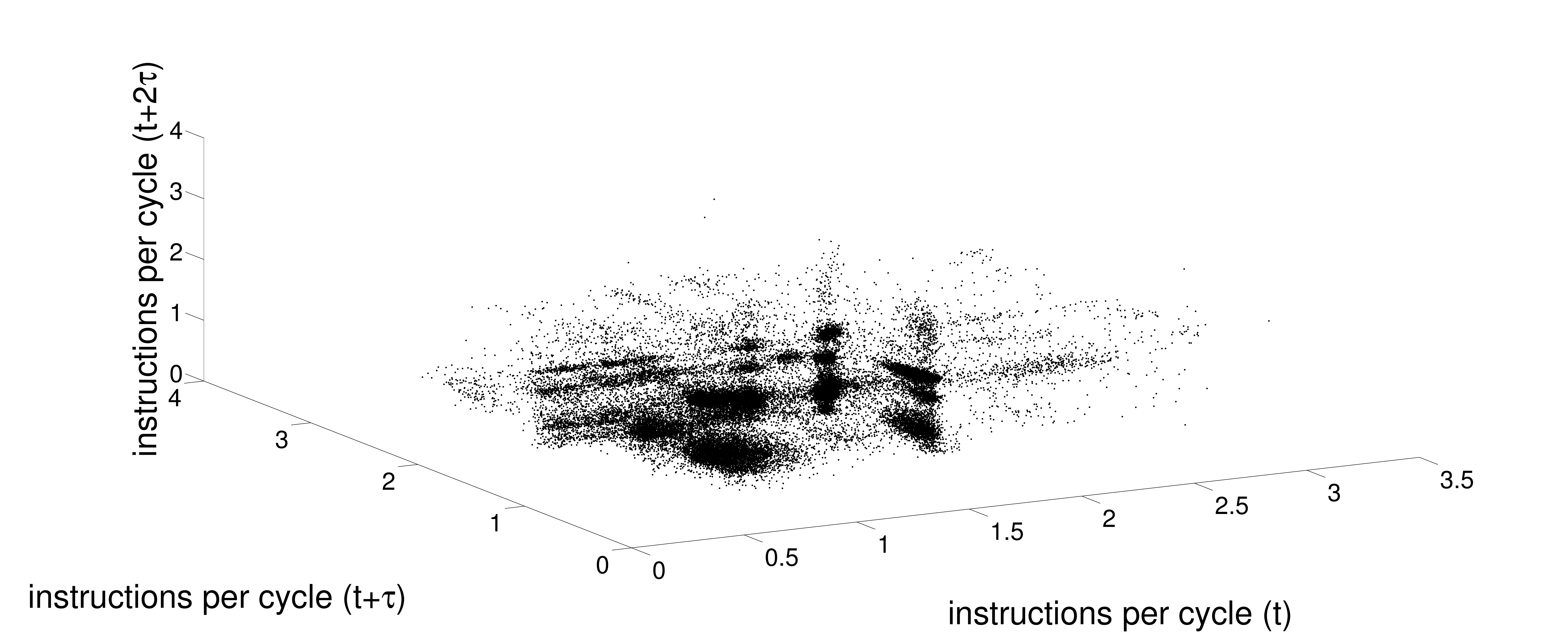}
  \includegraphics[width=.49\columnwidth]{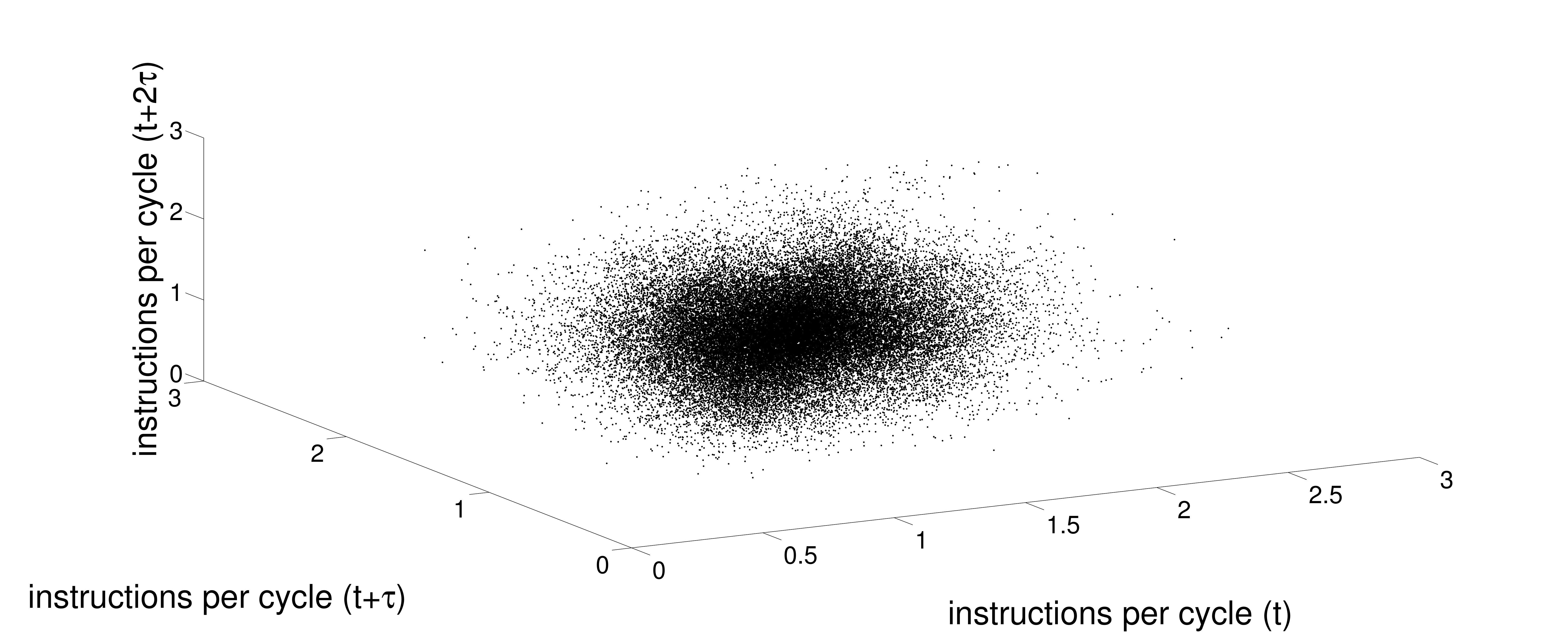}
    {\sphinx} \hspace{.4\columnwidth}{\gcc}
%
  \caption{3D projections of delay-coordinate embeddings of the traces
    from Figure~\ref{fig:ipctrace}.}
\label{fig:embedding}
\end{figure}

The first step in constructing a nonlinear forecast model of a
time-series data set like the ones in Figure~\ref{fig:ipctrace} is to
perform a delay-coordinate embedding using
equation~(\ref{eqn:takens}).  We followed standard procedures
\cite{kantz97} to choose appropriate values for the embedding
parameters: the first minimum of the mutual information
curve~\cite{fraser-swinney} as an estimate of the delay $\tau$ and the
false-nearest neighbors technique~\cite{KBA92}, with a threshold of
10-20\%, to estimate the embedding dimension $d_{embed}$.  For both
traces in 
Figure~\ref{fig:ipctrace}, $\tau=100000$ instructions and
$d_{embed}=12$.
A plot of the reconstructed dynamics of these 
two traces appears in Figure~\ref{fig:embedding}.
The coordinates of each point on these plots are differently delayed
elements of the IPC time series: that is, IPC at time $t$ on the first
axis, IPC at time $t+\tau$ on the second, IPC at time $t+2\tau$ on the
third, and so on.  An equivalent embedding (not shown here) of the
\row trace looks very like \gcc: a blob of points.  The embedded \col
dynamics, on the other hand, looks like {\sl two} blobs of points
because of its square-wave pattern.  Recall from
Section~\ref{sec:nonlinear} that these trajectories are guaranteed to
have the same topology as the true underlying dynamics, provided that
$\tau$ and $d_{embed}$ are chosen properly.  And structure in these
kinds of plots is an indication of determinism in that dynamics.

The nonlinear dynamics community has developed dozens of methods that
use the structure of these embeddings to create forecasts of the
dynamics; see~\cite{casdagli-eubank92,weigend-book} for overviews.
The Lorenz method of analogues (LMA) is one of the earliest and
simplest of these strategies~\cite{lorenz-analogues}.  LMA creates a
prediction of the future path of a point $\vec{x}_o$ through the
embedded space by simply finding its nearest neighbor and then using
{\sl that} point's future path as the forecast\footnote{The original
  version of this method requires that one have the true state-space
  trajectory, but others (e.g.,~\cite{kennel92}) have validated the
  theory and method for the kinds of embedded trajectories used
  here.}.  The nearest neighbor step obviously makes this algorithm
very sensitive to noise, especially in a nonlinear system.  One way to
mitigate that sensitivity is to find the $l$ nearest neighbors of
$\vec{x}_o$ and average their future paths.  These comparatively
simplistic methods work surprisingly well for computer-performance
prediction, as reported at IDA 2010~\cite{josh-ida2011}.  In the
following section, we compare the prediction accuracy of LMA models
with the MLR models of Section~\ref{sec:multilinear}.

\vspace*{-1mm}
\section{When and why are nonlinear models better at predicting computer performance?}
\label{sec:prediction}

\subsection{Procedure}
\label{sec:procedure}

Using the methods described in Sections~\ref{sec:multilinear}
and~\ref{sec:nonlinear}, respectively, we built and evaluated linear
and nonlinear models of performance traces from the four programs
described on page~\pageref{page:programs} (\gcc, \sphinx, \col and
\row), running on the computer described in Section~\ref{sec:methods}.
The procedure was as follows.  We held back the last $k$ points of
\label{page:horizon}
each time series (referred to as ``the test signal,'' $c_i$).
We then constructed the model with the remaining portion of the time
series (``the learning signal'') and used the model to build a
prediction $\hat{p}_i$.  We computed the root mean squared prediction
error between that prediction and the test signal in the usual way:
$$\err = \sqrt{\frac{\sum_{i=1}^k(c_i-\hat{p_i})^2}{k}}$$
To compare the results across signals with different units, we
normalized the \err as follows:
$$\nerr = \frac{\err}{max\{c_i\} - min\{c_i\}}$$ 
%
The smaller the \nerr, obviously, the more accurate the prediction.

\vspace*{-1mm}
\subsection{Results and Discussion}

First, we compared linear and nonlinear models of the two SPEC
benchmark programs: the traces in the top row of
Figure~\ref{fig:ipctrace}.  For \gcc, the nonlinear LMA model was
better than the linear MLR model (0.128 \nerr versus 0.153).  For
\sphinx, the situation was reversed: 0.137 \nerr for LMA and 0.116 for
MLR.  This was contrary to our expectations; we had anticipated that
the LMA models would work better because their ability to capture both
the gross and detailed structure of the trace would allow them to more
effectively track the regimes in the \sphinx signal.  Upon closer
examination, however, it appears that those regimes overlap in the IPC
range, which could negate that effectiveness.  Moreover, this
head-to-head comparison is not really fair.  Recall that MLR models
use {\sl multiple} measurements of the system---in this case, L2 total
cache misses, number of branches taken, and L2 instruction cache
misses---while LMA models are constructed from a {\sl single}
measurement (here, IPC).  In view of this, the fact that LMA beats MLR
for \gcc and is not too far behind it for \sphinx is impressive,
particularly given the complexity of these programs.  Finally, we
compared the linear and nonlinear model results to a simple ``predict
the mean'' strategy, which produces a 0.140 and 0.250 \nerr for \gcc
and \sphinx, respectively---higher than either MLR or LMA.

In order to explore the relationship between code complexity and model
performance, we then built and tested linear and nonlinear models of
the \row and \col traces.  The resulting \nerr values for these
programs, shown in the third and fourth row of
Table~\ref{tbl:results}, were lower than for \gcc and \sphinx,
supporting the intuition that simpler code has easier-to-predict
dynamics.
\begin{table}[htdp]
\begin{center}
\begin{tabular}{|c|c|c|c|c|}
\hline Program & Interrupt Rate (cycles) & LMA \nerr & MLR \nerr & naive \nerr  \\ \hline 
\gcc & 100,000 & 0.128&0.153&0.140 \\
\sphinx & 100,000 & 0.137&0.116&0.250 \\
\row &  100,000 & 0.063&  0.091& 0.078 \\
\col  & 100,000 &0.020 &0.032 & 0.045  \\ \hline 
\gcc & 1,000,000 & 0.196 &0.208&0.199 \\
\sphinx & 1,000,000 & 0.137&0.187&0.462\\
\row & 1,000,000 &0.057 &  0.129& 0.103\\
\col  & 1,000,000 &0.028 &  0.305& 0.312\\ \hline 
\end{tabular}
%
\caption{Normalized root-mean-squared error between true and predicted
  signals for linear (MLR), nonlinear (LMA), and ``predict the mean''
  forecast strategies}
\end{center}
\label{tbl:results}
\end{table}
Note that the nonlinear modeling strategy was more accurate than MLR
for {\sl both} of these simple four-line matrix initialization loops.
The repetitive nature of these loops leaves its signature in their
dynamics: structure that is exposed by the embedding process.  LMA
captures and uses that global structure---in effect, ``learning''
it---while MLR does not.  Again, LMA's success here is even more
impressive in view of the fact that the linear models require more
information to construct.  Finally, note that the LMA models beat the
naive strategy for both \row and \col, but the linear MLR model did
not.

Another important issue in modeling is sample rate.  We explored this
by changing the sampling rate of the traces while keeping the overall
length the same: i.e., by sampling the same runs of the same programs
at 1,000,000 instruction intervals, rather than every 100,000
instructions.  This affected the accuracy of the different models in
different ways, depending on the trace involved.  For \gcc, MLR was
still better than LMA, but not by as much.  For \sphinx, the previous
result (MLR better than LMA) was reversed.  For \row and \col, the
previous relationship not only persisted, but strengthened.  In both
of these traces, predictions made from MLR models were less accurate
than simply predicting the mean; LMA predictions were {\sl better}
than this naive strategy.  See the bottom four rows of Table~1 for a
side-by-side comparison of these results to the more sparsely sampled
results described in the previous paragraphs.

\begin{figure}
  \centering
  \includegraphics[width=0.8\columnwidth]{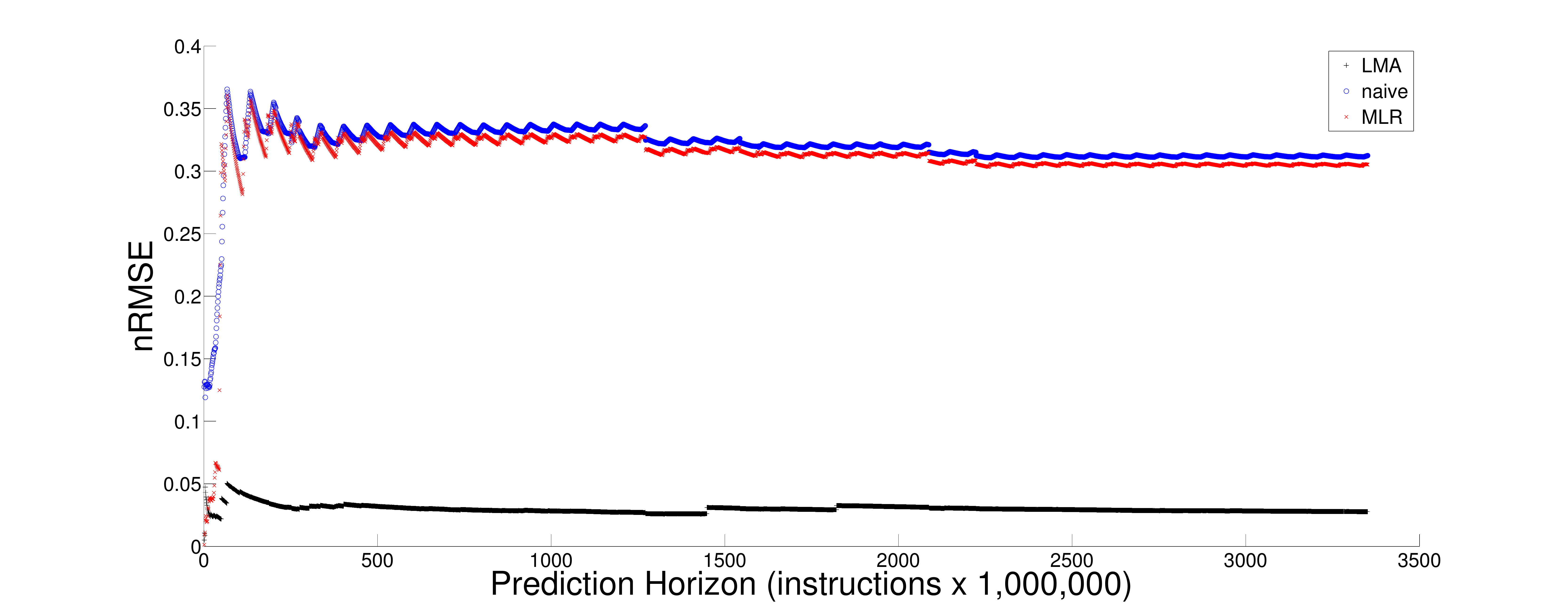}

%
%

  \includegraphics[width=0.8\columnwidth]{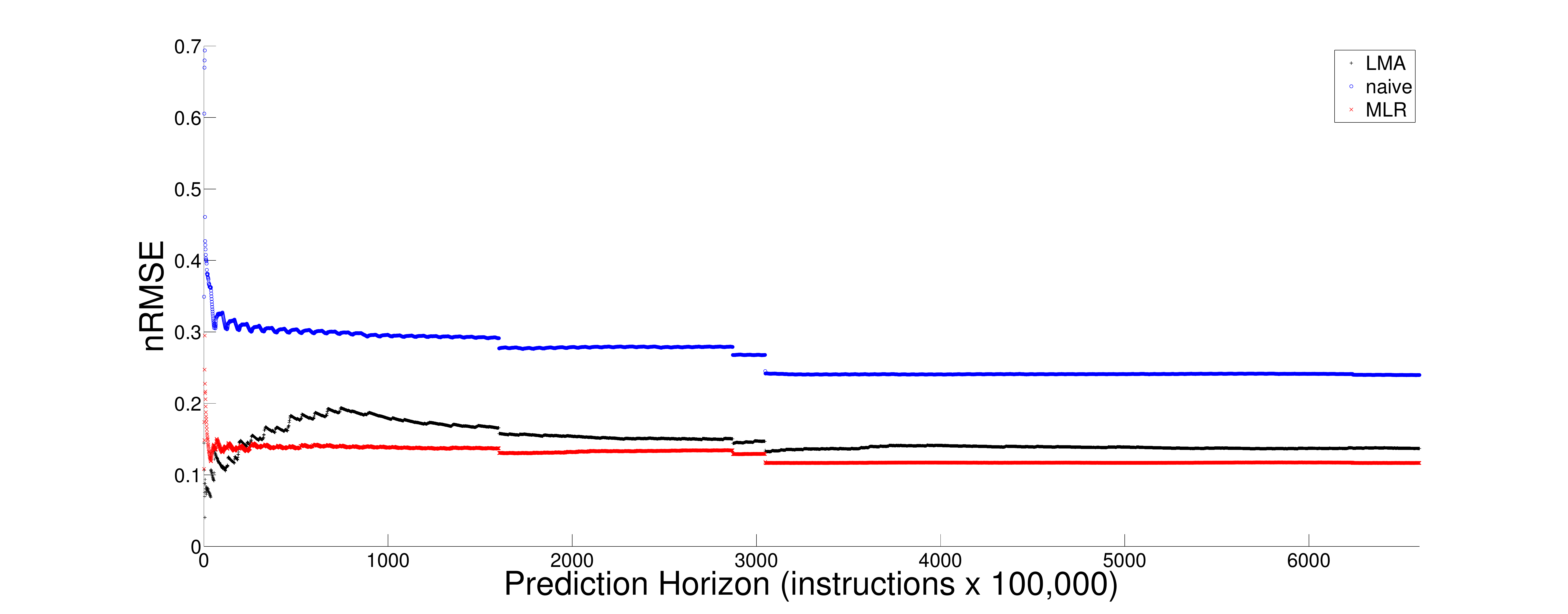}
%

  \caption{\nerr versus prediction horizon.Top: \col  Bottom: \sphinx}
  \label{fig:horizon}
\end{figure}

To explore which model worked better as the prediction horizon was
extended, we changed that value (the $k$ in
Section~\ref{sec:procedure}) and plotted \nerr.  In three of the four
traces---all but \sphinx---the nonlinear model held and even extended
its advantage as the prediction horizon lengthened; see
Figure~\ref{fig:horizon} for some representative plots.
The initial variability in these plots is an artifact of normalizing
over short signal lengths and should be disregarded.  The vertical
discontinuities (e.g., at the 1300 and 2100 marks on the horizontal
axis of the \col plot, as well as the 1600 and 3000 marks of \sphinx)
are also normalization artifacts\footnote{When the signal moves into a
  heretofore unvisited regime, that causes the $max - min$ term in the
  \nerr denominator to jump.}.  The sawtooth pattern in the top two
traces in \col nRMSE are due to the cyclical nature of that loop's
dynamics.  LMA captures this structure, while MLR and the naive
strategy do not, and thus produces a far better prediction.

\vspace*{-3mm}
\section{Conclusion}\label{sec:conclusion}
\vspace*{-2mm}

The experiments reported in this paper indicate that even a very basic
nonlinear model is generally more accurate than the state-of-the-art
linear model in the computer performance literature.  This result is
even more striking because those linear models require far more
information to build than the nonlinear models do---and they cannot be
used to predict further than one timestep into the future.  It is somewhat surprising that these linear models
work at all, actually, since many of the assumptions upon which they
rest are not satisfied in computer performance applications.
%
%
Nonlinear models that work in delay-coordinate embedding space may be
somewhat harder to construct, but they capture the structure of the
dynamics in a truly effective way.

%
It would be of obvious interest to apply some of the better linear
models that have been developed by the data analysis and modeling
communities to the problem of computer performance prediction.  Even a
segmented or piecewise version of multiple linear
regression~\cite{segmented-regression,piece-regression}, for instance,
would likely do a better job at handling the nonlinearity of the
underlying system.  The difficulty with that approach, of course, is
how to choose the breakpoints between the segments.  And MLR is not
really designed to be a temporal predictor anyway; linear predictors
like the ones presented in~\cite{LinearPredTutorial} might be much
more effective.  There are also regression-based {\sl nonlinear}
models that we could use, such as~\cite{ghkss}, as well as the many
other more-sophisticated models in the nonlinear dynamics
literature~\cite{casdagli-eubank92,weigend-book}.  It might be useful
to develop nonlinear models that use sliding windows in the data in
order to adapt to regime shifts, but the choice of window size is an
obvious issue.  Finally, nonlinear models that use multiple
probes---like MLR does---could be extremely useful, but the underlying
mathematics for that has not yet been developed.

%
\bibliographystyle{splncs03}
\bibliography{bib} 
\end{document}